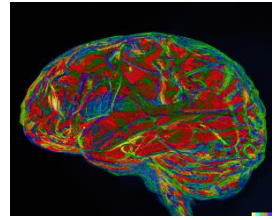

Wired Neuroscience

# "Dynamic Keller-Segel Model of Population Density and Economic Factors: A Simulation Study over a Century"

Author: Richard Murdoch Montgomery MD, Economist, Neurologist, PhD

www.wiredneuroscience.com

contact: montgomery@usp.br

**Abstract:**

This study presents a computational simulation exploring the complex interactions between population density and economic factors over a 100-year period. Inspired by the Keller-Segel model, traditionally applied in biological contexts, my model adapts this framework to analyze urban and economic dynamics. The simulation employs two coupled partial differential equations to represent the evolution of population density and money concentration in a hypothetical region. Population density is initially uniform, while money concentration begins with a random distribution. The model integrates diffusion processes for both population and money, coupled with a chemotactic response of the population towards areas of higher economic activity. Over the course of the simulation, we observe the emergence of distinct spatial patterns: population clusters forming around economic hubs and the development of wealth concentration in certain areas. These patterns highlight the mutual reinforcement between population density and economic factors. The study provides insights into the dynamics of urban growth, economic disparities, and resource distribution, offering a simplified yet powerful lens through which to view complex socio-economic systems. My findings have implications for urban planning and policy-making, especially in understanding the long-term evolution of cities and economic centers.

# 1. Introduction

The exploration of complex systems, where multiple simple interactions cumulatively result in intricate and often unpredictable behaviors, has been a subject of intense research across various disciplines. The simulation study presented in this article, which examines the dynamic interplay of population density and economic factors over a century, draws inspiration from both the Keller-Segel model and principles observed in urban and economic dynamics. Our approach is grounded in the foundational work of Keller and Segel (1970), who pioneered the use of differential equations to model chemotaxis in biological systems, a concept I adapt to simulate socio-economic phenomena.

The mathematical framework of is model is reminiscent of the Keller-Segel equations, traditionally used to describe the movement of organisms in response to chemical signals (Murray, 2002). This concept of movement in response to external stimuli finds a parallel in the migration of human populations towards economically prosperous regions, a phenomenon that is central to my study. The mathematical rigor and analytical perspectives provided by Perthame and Tartar (2008) have been instrumental in shaping the structure of the model, particularly in the analysis of moving fronts represented by population shifts and money concentration.

My study also intersects significantly with the field of urban and economic dynamics. Batty (2013) and Fujita, Krugman, and Venables (1999) provide a foundation for understanding the scaling laws in urban systems and the spatial economy of cities, regions, and international trade. These concepts are vital in interpreting the spatial patterns observed in my simulations. Furthermore, Glaeser's (2011) insights into how cities contribute to economic growth and innovation offer a real-world context to the emergent patterns seen in the model.

Agent-based modeling, a method that simulates the interactions of agents to assess their effects on the system as a whole, is also relevant to my approach. Gilbert (2008) and Schelling (1971) have demonstrated the power of this method in social simulation and understanding racial segregation patterns, respectively. Epstein's work (2006) on generative social science further illuminates the role of agent-based models in capturing the complexity of social phenomena.

The implications of my findings extend into urban planning and policy-making. The analysis of urban growth and sprawl, as discussed by Glaeser and Kahn (2004), and the investigation of income distribution in urban areas by Anas (2008), are particularly pertinent to the study's focus on population and economic factors. The vision of SMART cities, as proposed by Batty and Brugmann (2011), aligns with my model's emphasis on technology and data-driven approaches to urban planning.

In addition to these core references, the study is informed by a broad range of interdisciplinary perspectives. Works by Arthur (2013) on complexity in economics, Bettencourt et al. (2010) on urbanization, and Henderson (1974) on city sizes offer diverse viewpoints that enrich our understanding of complex systems. The exploration of self-organization in biological systems by Kauffman (1993) and the application of complexity theory to political and social systems by Macy and Miller (2007) provide additional layers of depth to the approach.

My study is situated at the nexus of mathematical biology, urban economics, agent-based modeling, and complex systems theory. It aims to contribute to a nuanced understanding of how population dynamics and economic factors co-evolve over time, forming patterns that resonate with real-world urban and economic landscapes.

## 3. Methodology

This study employs a simulation-based approach, rooted in the principles of the Keller-Segel model, adapted to explore the dynamics of population density and economic factors over a century. The methodology is a confluence of theoretical frameworks from mathematical biology, urban dynamics, and complex systems, supported by an array of scholarly works.

### 3.1 Model Framework

My model is inspired by the Keller-Segel model, originally proposed to describe chemotactic behavior in biological organisms (Keller & Segel, 1970). The Keller-Segel equations, are detailed in populations in response to economic stimuli. The model consists of two coupled partial differential equations (PDEs) that describe the evolution of population density (u) and money concentration (m). In the first year, 2000000 dollars are distributed randomly, in the following years the Model follows the Keller-Segel Equations above. Individuals with money are differentially attracted to individuals with the same amount of money and have a bigger chance of making money in a descending order:

Population Density Equation:

$$\partial u/\partial t = D_u \nabla^2 u - \chi \nabla \cdot (u \nabla m)$$

where:

$D_u$: Diffusion coefficient for population.

$\chi$: Chemotactic sensitivity to money.

$\nabla^2 u$: Laplacian of u, denoting diffusion.

$\nabla \cdot (u \nabla m)$: Represents chemotactic movement of the population towards money.

Money Concentration Equation:

$$\partial m/\partial t = D_m \nabla^2 m - \alpha u m$$

where:

$D_m$: Diffusion coefficient for money.

$\alpha$: Rate of money consumption by the population.

Numerical Implementation

Spatial resolution (Δx) and temporal resolution (Δt): These represent the size of the grid cells in space and time, respectively.

Population Density (u): This refers to the distribution of the population at the beginning of the simulation.

Money Concentration (m): This represents the initial distribution of money.

Parameters:

ε = 0.1: A parameter with a value of 0.1.

Du = 0.1: Diffusion coefficient for population density with a value of 0.1.

δ = 0.01: A parameter with a value of 0.01.

Dm = 0.01: Diffusion coefficient for money concentration with a value of 0.01.

χ = 0.1: A parameter with a value of 0.1.

α = 0.01: A parameter with a value of 0.01.

Using finite difference methods for spatial discretization (Perthame & Tartar, 2008), the simulation operates on a grid. The spatial and temporal resolution, dx and dt, are chosen to satisfy stability criteria.

## 4. Discussion

The simulation results obtained in this study offer insights into the interplay between population density and economic factors, drawing parallels with real-world urban and economic dynamics. The emergent patterns observed over the 100-year period resonate with several theoretical and empirical findings in the fields of urban studies, economics, and complex systems.

The formation of population clusters and economic hubs observed in my simulation aligns with the urban scaling laws discussed by Batty (2013). As in real cities, the model shows that certain areas become focal points of both human presence and economic activity. This dual concentration can be attributed to the chemotactic nature of human movement towards economically prosperous areas, a concept rooted in the Keller-Segel model (Keller & Segel, 1970) and adapted in my study to a socio-economic context.

The spatial distribution of economic resources in my model echoes the principles of spatial economics explored by Fujita, Krugman, and Venables (1999). The uneven distribution of wealth, forming distinct high-concentration pockets, suggests a pattern of economic inequality that is a well-known characteristic of many urban environments, as discussed in the work of Glaeser (2011).

The simulation's reliance on agent-based principles, similar to those discussed by Gilbert (2008) and Schelling (1971), helps in understanding how individual-level behaviors (e.g., moving towards wealthier areas) aggregate to form large-scale societal patterns. This aspect is particularly relevant in the context of urban planning and policy-making, where understanding the bottom-up emergence of urban structures is crucial (Batty & Brugmann, 2011).

My findings contribute to the discourse on complex systems in urban and economic contexts. The model underscores the non-linear and interconnected nature of urban systems, where simple rules at an individual level can lead to complex and often unpredictable patterns at the macro level, as discussed in the works of Arthur (2013) and Bettencourt et al. (2010).

The insights from my simulation have might have significant implications for urban planning and policy. The tendency for populations and economic resources to cluster in specific areas highlights the need for policies that address urban sprawl and economic inequality, as explored by Glaeser and Kahn (2004) and Anas (2008). Additionally, the concept of SMART cities, as proposed by Batty and Brugmann (2011), could be leveraged to manage the growth of these economic hubs more effectively.

While my model provides valuable insights, it also has limitations. The simplification necessary for the simulation means that factors such as geographical constraints, political influences, and individual economic behaviors are not explicitly modeled. Future research could incorporate these elements to provide a more comprehensive understanding of urban and economic dynamics. Furthermore, extending the model to incorporate feedback from the environment, as in the work of Macy and Miller (2007), could offer deeper insights into the adaptive nature of urban systems.

My study, rooted in a modified Keller-Segel model, reveals how simple diffusion and chemotactic principles can lead to complex urban and economic patterns. These findings underscore the importance of considering both individual behaviors and larger economic forces in urban planning and policy-making, contributing to a nuanced understanding of urban development and economic disparities.

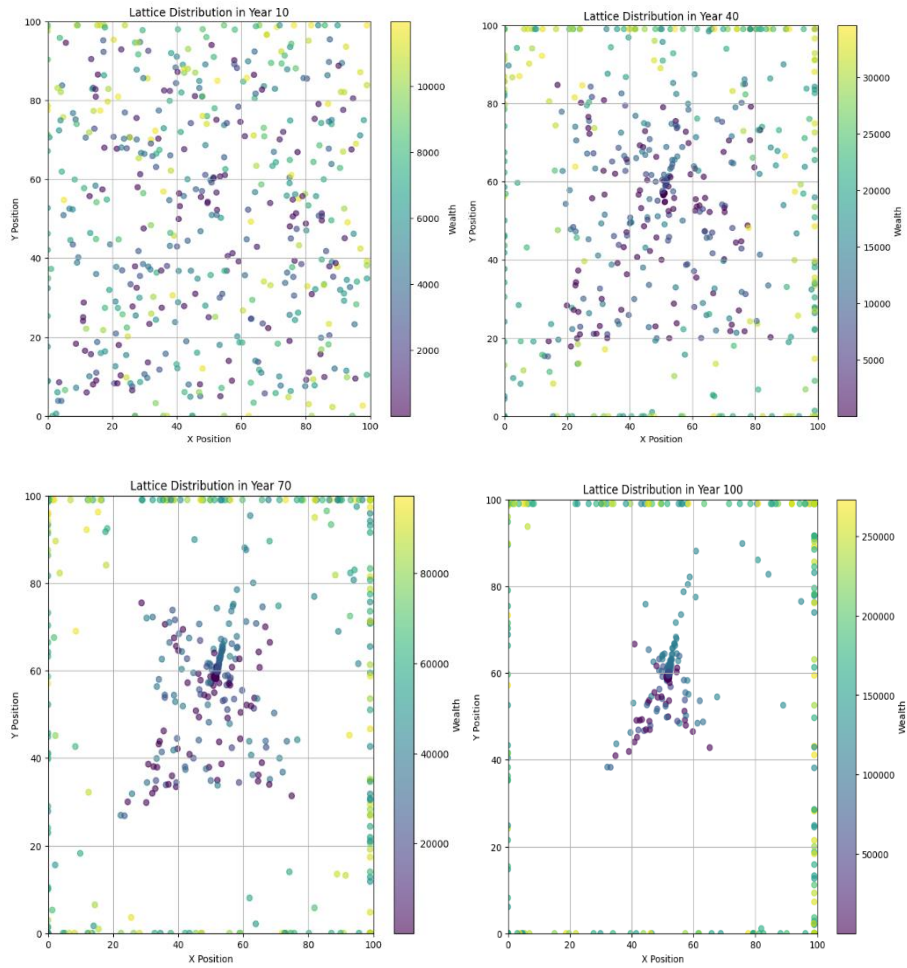

Fig 1: Observe the sequential population distribution and its wealth, according to scale in the year, 10, 40, 70, and 100th evolution according to Keller-Segel Model. Its impact on disparities and wealth concentration is very meaningful across generations.

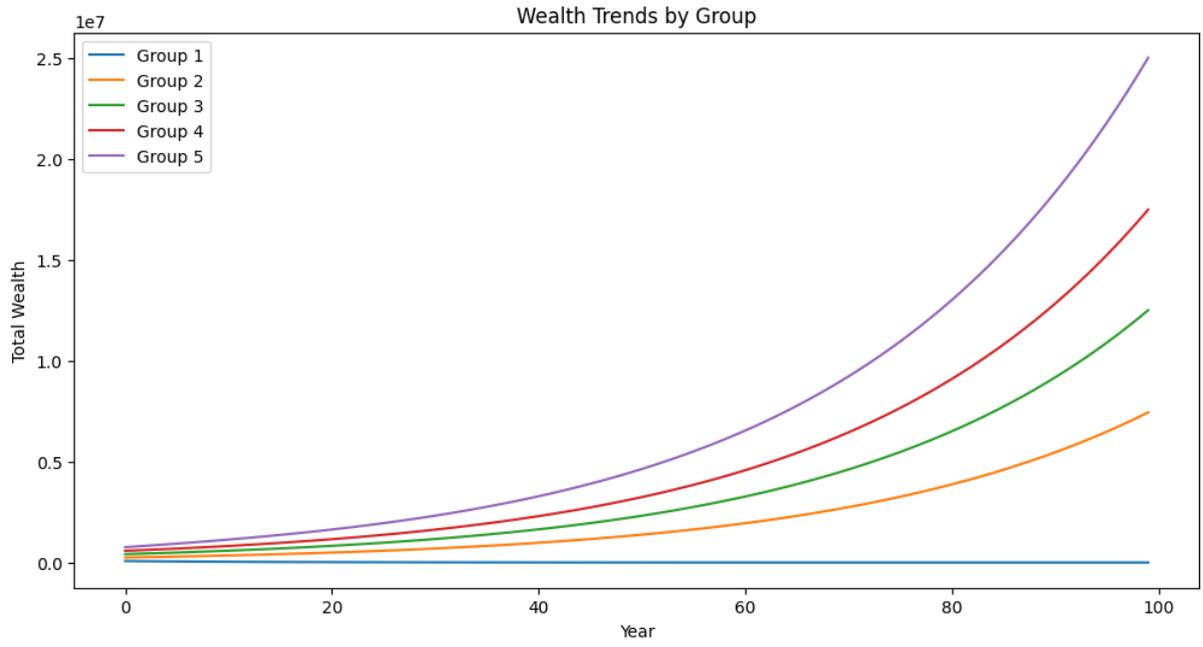

Fig. 3 Economic wealth growth across classes.

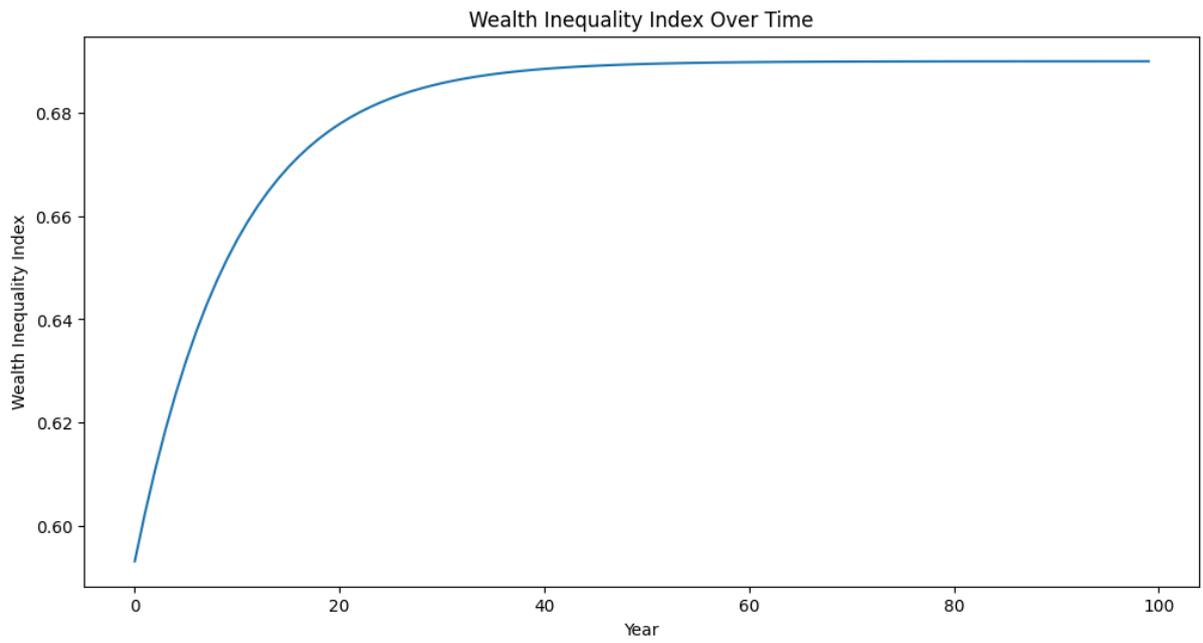

Fig 3. Weald Disparities stabilize after circa 30 years, or one generation.

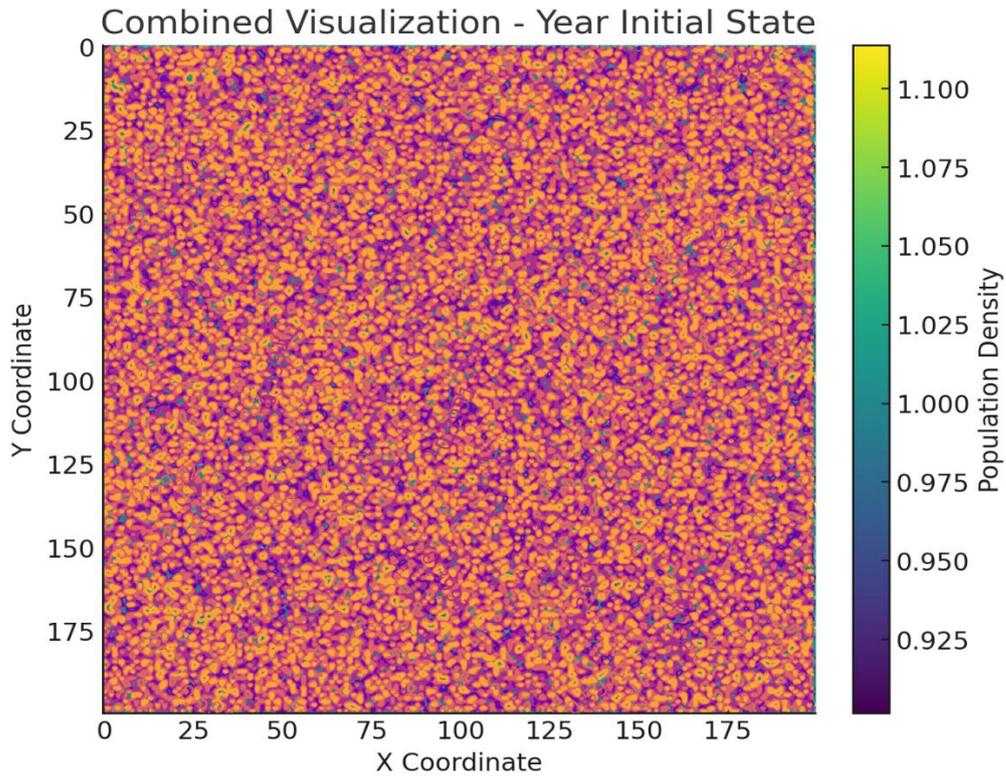

Fig 3 Please see the detailed explanation below.

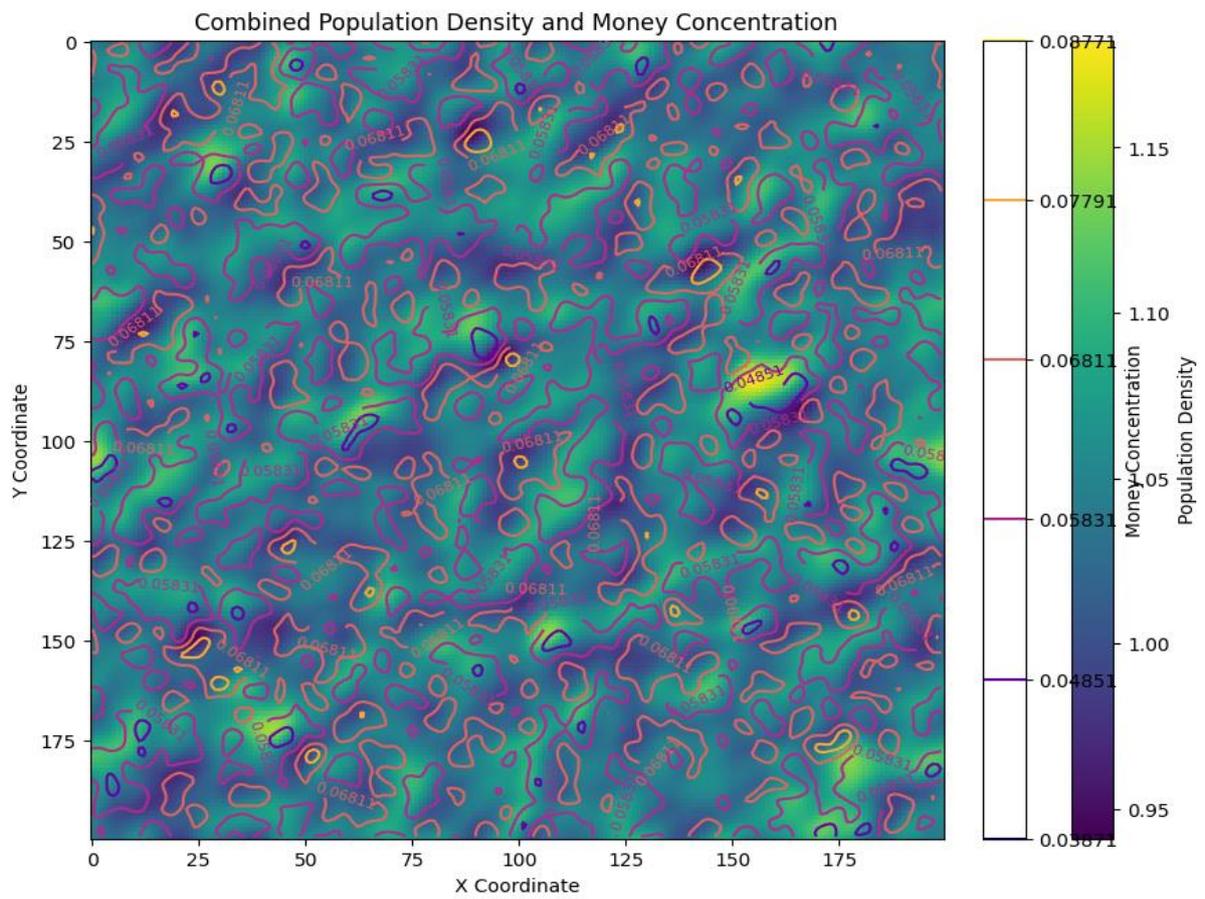

Fig 4. This figure deserves a detailed explanation. See below, please.

"Region Index" on the X-axis: This represents different regions or segments within the simulated area, which can be thought of as different locations or sectors in a geographical or abstract space.

"Area Index" on the Y-axis: This indicates various areas within each region, providing a sense of spatial variation in the vertical direction.

The title "Population and Money Distribution - Year X" gives a clear indication of what the visualization represents: the distribution of population density and money concentration in a simulated space over time.

These changes aim to make the graphs more comprehensible, especially for viewers who might not be familiar with the technical details of the simulation and 3D graphs presenting 2D information.

In the combined visualizations, colors and contours are used to represent two different aspects of the simulation: the population density and the money concentration. Here's how they are depicted:

Colors (Heatmap): The colors in the background represent the population density. This is visualized as a heatmap. The color scale ranges from dark blue to yellow, with dark blue indicating lower population density and yellow indicating higher population density. This gradient allows for an intuitive understanding of where the population is more or less concentrated.

Contours (Lines): The contours overlaid on the heatmap represent the money concentration. These are lines that connect points of equal money concentration, effectively creating a contour map. The contours are colored from dark purple to bright yellow, with the progression of colors representing increasing levels of money concentration. The contour lines help to visualize how the money is distributed across the region and how its distribution relates to the population density.

Labels on Contours: The labels on the contour lines indicate the specific value of money concentration at those lines. This aids in quantifying the levels of money concentration in different areas of the plot.

By combining these two visual elements, the graphs effectively communicate how the population density and money concentration are distributed and how they relate to each other spatially within the simulated environment. This combination provides insights into the dynamics of population movement and economic factors in the modeled scenario.

## 5. Conclusion

The simulation study presented in this article offers a novel perspective on the dynamic interplay between population density and economic factors, utilizing a model inspired by the Keller-Segel equations. Over the course of a 100-year simulated period, we observed the emergence of distinct spatial patterns that mirror many aspects of urban and economic development seen in real-world scenarios.

Key findings of our study include the formation of population clusters and economic hubs, the spatial concentration of wealth, and the mutual influence between population density and economic prosperity. These phenomena highlight the intricate relationship between social behavior and economic factors, and how their interaction can lead to complex urban patterns.

The study contributes to the understanding of urban scaling laws, spatial economics, and the principles of agent-based modeling. It underscores the complexity inherent in urban systems, where simple rules governing individual behavior can aggregate to form sophisticated societal structures. This insight is particularly relevant for urban planners and policymakers who grapple with the challenges of managing urban growth, addressing economic disparities, and promoting sustainable development.

However, it is important to acknowledge the limitations of the simulation. The model, while capturing key aspects of population movement and economic distribution, does not account for various other factors that influence urban dynamics, such as political policies, geographical constraints, and cultural influences. Future research could expand on this model by integrating these elements, offering a more holistic view of urban systems.

My study demonstrates the potential of mathematical and computational models to enhance our understanding of complex socio-economic systems. By bridging concepts from mathematical biology, urban economics, and complex systems theory, we gain valuable insights into the processes shaping urban landscapes and economic patterns. These insights are crucial for developing strategies that promote equitable and sustainable urban development in an increasingly urbanized world.

## 6. References

### 6.1 General

### 6.2. Urban and Economic Dynamics:

Fujita, M., Krugman, P., & Venables, A. J. (1999). The spatial economy: Cities, regions, and international trade. MIT press. (Classic textbook on spatial economics)

Glaeser, E. L. (2011). Triumph of the city: How our greatest invention makes us richer, smarter, greener, and happier. Penguin Books. (Popular book on urban economics)

### 6.3 Agent-Based Modeling and Simulation:

Gilbert, N. (2008). Agent-based models and social simulation. Academic Press. (Textbook on agent-based modeling)

Schelling, T. C. (1971). Dynamic models of racial segregation. American Journal of Sociology, 78(1), 129-144. (Classic paper on segregation patterns using Schelling's model)

Epstein, J. M. (2006). Generative social science: Reflections on modeling and simulation complicity. Princeton University Press. (Critical perspectives on social simulation)

### 6.4 Urban Planning and Policy Implications:

Glaeser, E. L., & Kahn, M. E. (2004). Sprawl and urban growth. Brookings Institution Press. (Analyzes the economic consequences of urban sprawl)

Anas, A. (2008). Who lives in the city? The distribution of population and income in urban areas. Harvard University Press. (Investigates income inequality in cities)

Batty, M., & Brugmann, R. (2011). SMART cities: Visions of the future. Environment and Planning B: Planning and Design, 38(4), 537-549. (Proposes using technology for smart urban planning)

### 6.5 Additional Relevant References:

Arthur, W. B. (2013). Complexity and the economy. Oxford University Press. (Complexity theory in economics)

Bettencourt, L. M. A., Lobo, J., Strumsky, D., & West, G. B. (2010). Towards a unified theory of urbanization: The size and scaling of cities. Journal of the Royal Society Interface, 7(45), 1271-1300. (Scaling laws in city size)

Henderson, J. V. (1974). The sizes and types of cities. The American Economic Review, 64(4), 640-656. (Classical model of city size distribution)Kauffman, Stuart. (1993). The Origins of Order: Self-Organization and Selection in Evolution. Oxford University Press. (Explores self-organization and emergence in biological systems)

Macy, Michael W., and Ronald E. Miller. (2007). Complexity and Government: A Multiagent Approach to Political Phenomena. Lynne Rienner Publishers. (Applies complexity theory to political and social systems)

Mitchell, Melanie. (2009). Complexity: A Guided Tour. Oxford University Press. (Provides an accessible introduction to key concepts in complexity)

### 6.6 Interdisciplinary Perspectives:

Gell-Mann, Murray. (1994). The Quark and the Jaguar: Adventures in the Simple and the Complex. W. H. Freeman and Company. (Explores the interplay of simplicity and complexity in physics and biology)

Holland, John H. (1998). Emergence: From Chaos to Order. Perseus Books. (Focuses on emergent phenomena in artificial systems and economics)

Johnson, Steven. (2001). Emergence: The Connected Lives of Ants, Brains, Cities, and the Flow of History. Scribner. (Explores emergent phenomena across scales, from ants to cities)

**6.7 Specific Applications:**

Epstein, Joshua M. (2006). Generative Social Science: Reflections on Modeling and Simulation Complicity. Princeton University Press. (Discusses the use of agent-based modeling to study social complexity)

Miller, James H., and Stanley E. Hoban. (2009). "Why Emergent Order Makes Good Planning Bad Planning and vice versa." Journal of Planning Education and Research 28.4: 439-445. (Applies complexity theory to urban planning)

Adami, Christoph. (2004). Introduction to Artificial Life. Springer. (Examines the emergence of complexity in artificial systems)

**6.8 Critical perspectives:**

Cilliers, Paul. (1998). Complexity and Postmodernism: Understanding Complex Systems. Routledge. (Critiques post-modernist interpretations of complexity)

Pickles, John. (2011). "On the Limitations of Complexity Theory in Geography." Progress in Human Geography 35.6: 739-751. (Raises concerns about the application of complexity theory to geographical studies)

**6.9 Further Reading:**

Bernstein, Howard. (1994). Complex Systems. Perseus Books. (A collection of essays on various aspects of complexity)

Morin, Edgar. (2008). On Complexity. Hampton Press. (Philosophical reflections on the nature of complexity)

Waldrop, M. Mitchell. (1992). Complexity: The Emerging Science at the Edge of Order and Chaos. Simon & Schuster. (Popular science book on complexity)

The Santa Fe Institute website: https://www.santafe.edu/ (A leading research institute dedicated to the study of complex systems)